\documentclass[12pt]{article}

\catcode`\@=11

\global\arraycolsep=2pt
\usepackage{amsbsy,amssymb,latexsym,amsfonts,amsmath}
\usepackage{graphicx,color}

\allowdisplaybreaks

\begin{document}

\begin{flushright}
\parbox{4.2cm}
{RUP-22-9}
\end{flushright}

\vspace*{0.7cm}

\begin{center}
{ \Large New conformal field theories by gauging electric and magnetic currents}
\vspace*{1.5cm}\\
{Yu Nakayama}
\end{center}
\vspace*{1.0cm}
\begin{center}

Department of Physics, Rikkyo University, Toshima, Tokyo 171-8501, Japan

\vspace{3.8cm}
\end{center}

\begin{abstract}
Old folklore says that there is no non-trivial renormalization group fixed point with $U(1)$ gauge symmetry in four dimensions, but it can be circumvented by the existence of magnetic monopoles. We propose to construct (potentially infinitely many) novel types of field theories with $U(1)$ gauge symmetry by gauging $U(1) \times U(1)$ global symmetry electrically and magnetically. If the construction is consistent, the resulting theory will most likely possess a non-trivial renormalization group fixed point for the $U(1)$ gauge coupling constant and it will be a non-trivial conformal field theory without the  $U(1)\times U(1)$ global symmetry in the infrared upon tunings of other relevant deformations. If the construction is CP-violating, a renormalization of the $\theta$-parameter is accompanied.
\end{abstract}

\thispagestyle{empty} 

\setcounter{page}{0}

\newpage

\section{Introduction}
Old folklore says that there is no non-trivial renormalization group fixed point with $U(1)$ gauge symmetry in four dimensions. The simple argument goes like this (see e.g. \cite{Adler:1972in}): assume that the field strength is determined by a gauge potential: $F_{\mu\nu} = \partial_\mu A_\nu - \partial_\nu A_\mu$ and study the spectral representation 
\begin{align}
& \int d^4x e^{-ipx} \langle 0|T F_{\mu\nu}(x) F_{\rho\sigma}(0)|0\rangle \cr
&= -i(\eta_{\mu\rho} p_\nu p_\sigma - \eta_{\nu\rho} p_\mu p_{\sigma} - \eta_{\mu\sigma} p_\nu p_{\rho} +\eta_{\nu\sigma} p_\mu p_{\rho}) \int ds \frac{\rho(s)}{s+p^2} \ ,
\end{align}
given by the electric spectral density $\rho(s)\ge 0$. The wavefunction renormalization for $A_\mu$ is therefore given by
\begin{align}
Z_3 = \int ds \frac{\rho(s)}{1+\frac{s}{\mu^2}} , 
\end{align}
but it is monotonically increasing (because $\rho\ge 0$) with respect to $\log \mu$. Thus we find
\begin{align}
\beta_e = \frac{1}{2} e \frac{\partial \log Z_3}{\partial \log \mu} \ge 0 \ . 
\end{align}
If there were any renormalization group fixed point for $e$, then the spectral density $\rho$ should be zero above (or below) a certain scale, and the two-point function of $F_{\mu\nu}$ would approach a free theory expression, implying that the fixed point must be trivial.

If we assume conformal invariance at the fixed point, we could offer a more direct argument \cite{Argyres:1995xn}. The Bianchi identity suggests $\partial^\mu \tilde{F}_{\mu \nu} =0$, and the unitarity of the conformal algebra says that the conformal primary operator $F_{\mu\nu}$ must have a canonical dimension $\Delta_{F}=2$. Then applying the unitarity argument again  to the Maxwell equation $\partial_\mu F^{\mu\nu} = J_e^\nu$ leads to the conclusion $J^\mu_e =0$ because the operator with the canonical dimension must be a singular vector (i.e. satisfying a free equation of motion). Consequently, the fixed point must be trivial. 

The above argument can be circumvented by the existence of magnetic monopoles. The spectral representation must be modified due to the fact that $F_{\mu\nu}$ cannot be expressed by a single gauge potential $A_\mu$. Accordingly the Bianchi identity  no longer holds, and instead we have the dual Maxwell equation $\partial_\mu \tilde{F}^{\mu \nu} = J_m^\nu$. Then a non-trivial conformal fixed point with  non-zero anomalous dimensions is possible.

In this paper, we propose to construct (potentially infinitely many) novel types of field theories with $U(1)$ gauge symmetry by gauging $U(1) \times U(1)$ global symmetry electrically and magnetically. If the construction is consistent, the resulting theory will most likely possess a non-trivial renormalization group fixed point for the $U(1)$ gauge coupling constant and it will be a non-trivial conformal field theory without the $U(1)\times U(1)$ global symmetry in the infrared upon tunings of other relevant deformations. If the construction is CP-violating, renormalization of the $\theta$-parameter is accompanied.

\section{Construction}
Our starting point is a four-dimnsional conformal field theory with $U(1)_e\times U(1)_m$ global symmetry. Correspondingly, the theory has two conserved currents $J^\mu_e$ and $J^\mu_m$. We are going to gauge $J^\mu_e$ electrically and $J^\mu_m$ magnetically with respect to one physical gauge field (i.e. with only two polarization degrees of freedom). For the success of the gauging, we assume that the charges associated with $J^\mu_e$ and $J^\mu_m$ (i.e. charges of all the local operator of the conformal field theory) are quantized \cite{Dirac:1931kp}\cite{Zwanziger:1968rs}:
\begin{align}
e_i g_j - e_j g_i = \frac{n_{ij}}{2} \ , \label{Dirac}
\end{align}
where $n_{ij}$ are integers.
For example, this is satisfied if the conformal field theory is given by two copies of the same $U(1)$ symmetric theories with integral charges (e.g. a free complex scalar or a free Dirac fermion).
We will henceforth assume that $J^\mu_e$ and $J^\mu_m$ are normalized in this manner. 

We need some information from the underlying conformal field theories. First of all, we define the so-called current central charges\footnote{Our normalization is $k_{ee}$ for a comlpex scalar with charge one is given by $1/4$ and that for a Weyl fermion with charge one is given by $1/2$.} 
\begin{align}
\langle J^\mu_e(x) J^\nu_e(0) \rangle &= \frac{k_{ee}}{x^6} I^{\mu\nu(x)} \cr
\langle J^\mu_e(x) J^\nu_m(0) \rangle &= \frac{k_{em}}{x^6} I^{\mu\nu(x)} \cr
\langle J^\mu_m(x) J^\nu_m(0) \rangle &= \frac{k_{mm}}{x^6} I^{\mu\nu(x)} \ ,
\end{align}
where $I^{\mu\nu}(x) = \eta^{\mu\nu} - \frac{x^\mu x^\nu}{x^2}$. 
We assume the current central charge matrix  is a positive matrix. When it is degenerate, we effectively have only one $U(1)$ symmetry (or the theory is non-unitarity), and the resulting construction will not be interesting.

When the theory preserves $\mathrm{CP}$, it is customary to make $J^\mu_e$ and $J^\mu_m$ possess opposite $\mathrm{CP}$ charges (e.g. $\mathrm{CP}(J_e) = +1$ and $\mathrm{CP}(J_m)=-1$). In such cases, we should necessarily have $k_{em}=0$. Otherwise, the assignment of this CP charge is inconsistent (but we might be able to change the definition of $\mathrm{CP}$ more non-trivially).

We note that the charge quantization condition \eqref{Dirac} does not fix the normalization of $J^\mu_e$ and $J^\mu_m$ completely. For example, one may simply multiply integers and still they satisfy \eqref{Dirac}, which results in different $k_{ee}$, $k_{em}$ and $k_{mm}$. A different choice of the normalizations of $J^\mu_e$ and $J^\mu_m$ will sometimes give a different theory after the gauging. 

The other necessary condition is the absence of the anomaly \cite{Csaki:2010rv}. We assume that the  't Hooft anomaly coefficient (that can be read from the three-point functions among $J^\mu_e$ and $J^\mu_m$) all vanish: $C_{eee} = C_{eem} = C_{emm} = C_{mmm} = 0$. Otherwise, the gauging will become inconsistent. Note that even if we will eventually introduce only one gauge degree of freedom, the anomaly condition is as if we were gauging $U(1)_e \times U(1)_m$ separately by two gauge fields.

At this point, one may simply declare that we are going to gauge $J^\mu_e$ electrically and $J^\mu_m$ magnetically and add the gauge kinetic term and the $\theta$-terms. To be more concrete, we use Zwanziger's formulation to write down the explicit action \cite{Zwanziger:1970hk}. For this purpose, we introduce two independent vector fields $A_\mu$ and $B_\mu$ and an arbitrary (space-like) four-vector $n^\mu$ that corresponds to a direction of Dirac strings to be attached.\footnote{{The similar structure appears in two-dimensions, where in order to make the momentum symmetry and winding symmetry manifest, we need to double the apparent degrees of freedom while breaking the Lorentz symmetry \cite{Tseytlin:1990va}.}}

We will use the notation used in \cite{Csaki:2010rv}\cite{Zwanziger:1970hk}: 
\begin{align}
a \cdot b & = a^\mu b_\mu \cr 
(* F)^{\mu\nu} & = \frac{1}{2}{\epsilon^{\mu\nu \alpha\beta}} F_{\alpha\beta} \cr
(a\wedge b)^{\mu\nu} &= a^\mu b^\nu - b^\nu a^\mu \cr
(a\cdot *(b\wedge c))^\nu &= \epsilon^{\mu\nu\alpha\beta} a_{\mu} b_{\alpha}c_\beta \ . 
\end{align}

The gauge part of the Lagrangian is given by 
\begin{align}
L_{\mathrm{gauge}} =& \mathrm{Im} \frac{\tau}{8\pi n^2} \left([n \cdot \partial \wedge  (A+iB)] \cdot [n \cdot \partial \wedge (A-iB)] \right) \cr
&-\mathrm{Re}  \frac{\tau}{8\pi n^2} \left([n \cdot \partial \wedge  (A+iB)] \cdot [n \cdot *\partial \wedge (A-iB)] \right) \ , 
\end{align}
where $\tau = \frac{\theta}{2\pi} + \frac{4\pi i}{e^2}$ is the complexified coupling constant. Note that we have doubled the vectorial degrees of freedom, but it is halved by the projection on to $n^\mu$ direction. 

We are going to gauge a conformal field theory with global $U(1)_e\times U(1)_m$ symmetry by adding the ``local" interaction 
\begin{align}
L_{\mathrm{gauging}} = \mathrm{Re}[(A-iB)\cdot(J_e+\tau J_m)] \ ,
\end{align}
where the Witten effect \cite{Witten:1979ey} has been incorporated when $\theta \neq 0$ \cite{Csaki:2010rv}. 
If we vary the total action with respect to $A_\mu$ and $B_\mu$, we obtain the electric as well as magnetic Maxwell equations:
\begin{align}
\frac{\mathrm{Im}(\tau)}{4\pi} \partial_\mu(F^{\mu\nu} + i *F^{\mu\nu}) = J_e^\nu + \tau J_m^\nu \ ,  \label{eom1}
\end{align}
Here we define the field strength $F_{\mu\nu}$ by
\begin{align}
n^2 F_{\mu\nu} = \left( n \wedge [n \cdot (\partial \wedge A)] \right)_{\mu\nu}  -  *\left(n \wedge [n\cdot (\partial \wedge B)] \right)_{\mu\nu} , 
\end{align}
which must be supplemented by current conservations
\begin{align}
\partial_\mu J^\mu_e = \partial_\mu J^\mu_m = 0 \ . \label{eom2}
\end{align}
The current conservations are also necessary in order to assure that the formalism is invariant under two gauge transformations $A_\mu \to A_\mu + \partial_\mu \Lambda_{e}$ and $B_\mu \to B_\mu + \partial_\mu \Lambda_{m}$.

We note that while the action is not manifestly invariant under the Lorentz transformation due to the presence of $n^\mu$, the equations of motion look like Lorentz invariant. 
We also note that \eqref{eom1} and \eqref{eom2} are also (classically) conformal invariant if we assume $F_{\mu\nu}$, $J_e^\mu$ and $J_m^\mu$ transform as primary fields with no anomalous dimensions.\footnote{To avoid the confusion, the assumption that there is no anomalous dimension here is not quantum mechanically correct. What happens at the non-trivial conformal fixed point will be discussed in the next section.} 

We may show that the above action is $SL(2,\mathbb{Z})$ invariant (i.e. $a,b,c,d \in \mathbb{Z}, ad-bc=1$):
\begin{align}
\tau' &= \frac{a\tau + b}{c\tau +d} \cr 
{J'}_e^\mu &= -b J_m^\mu + a J_e^\mu \cr
{J'}_m^\mu &=  d J_m^\mu - c J_e^\mu \cr 
A' +  iB' & = (c\tau^* + d) (A+iB) \ ,  \label{s1}
\end{align}
under which
\begin{align}
k'_{ee} &= a^2 k_{ee} - 2ba k_{em} + b^2 k_{mm} \cr
k'_{em} & = -ac k_{ee} +(bc +ad) k_{em} -bd k_{mm} \cr
k'_{mm} & = c^2 k_{ee} -2d c k_{em} + d^2 k_{mm} \ . \label{s2}
\end{align}

\section{Properties and Exmaples}
The first question asked by Zwanziger was if the electrically and magnetically gauged theory thus constructed possesses the Lorentz symmetry \cite{Zwanziger:1970hk}\cite{Brandt:1978wc}. The equations of motion look Lorentz invariant, but it may not be sufficient. There, the conserved energy-momentum tensor was constructed but it was not symmetric. The non-symmetric part is characterized by a mildly non-local term:
\begin{align}
T^{\mu\nu} = T^{\mu\nu}_{\mathrm{Maxwell}} 
-n^\mu \left( n \cdot *[(n\cdot \partial)^{-1} J_e \wedge (n\cdot \partial)^{-1} J_m] \right)^\nu \ . \label{unwanted}
\end{align}

The physical meaning of this term is as follows. Suppose $J^\mu_m$ is generated by a point-like magnetic monopole. To describe motion of an electrically charged particle in this background, we need to introduce  a Dirac string in $n^\mu$ direction from the monopole. The extra term \eqref{unwanted} is zero unless the electrically charged particle crosses the Dirac string. The entire point is if such configurations can be avoided by consistently changing the direction of Dirac strings. 

The invisibility of the Dirac string is directly related to the Dirac charge quantization condition \cite{Brandt:1978wc}, and the corresponding statement also holds in quantum field theories. In other words, as long as the Dirac charge quantization is satisfied we may effectively think  as if $(n\cdot \partial)^{-1} J_e \wedge (n\cdot \partial)^{-1} J_m  = 0$ with Lorentz invariance. 

At the classical level, we observe that the trace of \eqref{unwanted} vanishes. Thus, if the matter action is conformal invariant, the entire energy-momentum tensor is traceless and it is scale invariant. In order to show the enhanced symmetry of conformal invariance, we need to show that the energy-momentum tensor is traceless and symmetric. As we have discussed above, if we neglect the extra term in  \eqref{unwanted}, then we may say that the gauged theory is classically conformal invariant. 

Quantum mechanically, the coupling constant $e$ and $\theta$ are renormalized. The ``one-loop" universal part of the beta function is fixed by conformal data of the ungauged conformal field theories: 
\begin{align}
\beta_{\tau} = \frac{\partial \tau}{\partial \log \mu} =  -\frac{i}{16\pi^2}(k_{ee} + 2k_{em} \tau + k_{mm} \tau^2) \ ,  \label{betaf}
\end{align}
or
\begin{align}
\beta_e &= \frac{\partial e}{\partial \log\mu} = \frac{e^3}{12\pi^2} \left( k_{ee} + \frac{\theta}{2\pi} 2k_{em} + \left(\frac{\theta}{2\pi}\right)^2 k_{mm} - k_{mm} \frac{16\pi^2}{e^4} \right) \ \cr
\beta_\theta & = \frac{\partial \theta}{\partial \log\mu} = \frac{1}{e^2} \left(k_{em} + \frac{\theta}{2\pi}k_{mm} \right) \ .  
\end{align}
The philosophy to obtain this beta function is the same as in  \cite{Argyres:1995jj}\cite{Csaki:2010rv} (see also \cite{Laperashvili:1999pu}), where we obtain the one-loop beta function in one duality frame (e.g. in $k_{em}=k_{mm}=0$) and then rotate back and add each contributions.
Unfortunately, this one-loop formula is based on a potentially uncontrolled approximation because $e$ and $e^{-1}$ cannot be simultaneously small in an arbitrary manner, but we expect that the general feature of the renormalization group flow can be studied.\footnote{It would be interesting if the introduction of supersymmetry makes this formula exact,}

Let us look for a renormalization group fixed point of $e$ and $\theta$. It is equivalent to find a  scale invariant fixed point. Let us recall that when the trace of the energy-momentum tensor is given by a divergence of some vector operators, it is scale invariant, if it vanishes up to improvement, then it is conformal invariant. See \cite{Nakayama:2013is} for a review.

In our case, one may wonder if we can relax the condition $\beta_\theta = 0$ to find a scale invariant (but not conformal) fixed point. Indeed, if we were working in electrically gauged theory, the $\theta$-term would couple to a divergence of Chern-Simons current so we would not have to set $\beta_\theta=0$ to find a scale invariant fixed point. However, here due to the presence of the magnetic current, the  $\theta$-term is no longer a total derivative and we have to demand $\beta_\theta=0$. 

If there is an additional anomalous current whose divergence is given by the $\theta$-term (i.e. $\partial_\mu J_A^\mu = c F_{\mu\nu} \tilde{F}^{\mu\nu}$),\footnote{For example, one may simply add massless electrically charged Dirac fermions.} then we do not have to impose $\beta_\theta =0$ to obtain scale (but not conformal) invariant fixed point. However, as we can see from the structure of the beta function, $\beta_e=0$ line is not invariant under the renormalization group flow of $\theta$ unless $\beta_\theta=0$, so we will necessarily end up with the conformal invariant fixed point even in this case. This agrees with the general expectation that scale invariant field theories are conformal invariant in four dimensions \cite{Luty:2012ww}\cite{Dymarsky:2013pqa}.

The zero of the beta functions (within the ``one-loop" approximation) is located at
\begin{align}
\tau_* = \frac{-k_{em} \pm i \sqrt{ k_{ee} k_{mm}- k_{em}^2}}{k_{mm}} \ ,
\end{align}
or
\begin{align}
\frac{\theta_*}{2\pi} & = -\frac{k_{em}}{k_{mm}} \cr
e_*^4  &= \frac{ 16 \pi^2 k_{mm}^2}{k_{ee} k_{mm} - k_{em}^2} \ . 
\end{align}
We have assumed $k_{ee} k_{mm} - k_{em}^2 > 0$. Otherwise, the ungauged theory would have been non-unitary (and the gauged theory would be non-unitary as well).

Note that the fixed point is infrared attractive assuming that there are no other relevant deformations that come from the conformal field theories we started with.   As we have already mentioned, given a $U(1)_e\times U(1)_m$ symmetry, the choices of $k_{ee}$, $k_{em}$ and $k_{mm}$ are more or less arbitrary up to charge quantization condition. In particular, we can take large $k_{ee}$, $k_{em}$ and $k_{mm}$ limit (e.g. one electron with electric charge $100$ and one monopole with  magnetic charge $101$, or one electron with electric charge $1000$ and one monopole with magnetic charge $1001$.) so that we may have infinitely many non-trivial fixed points with effectively continuous fixed values of $\theta_*$ and $e_*$ (while operator contents are more or less the same). 

{There exist duality invariant fixed points in these beta functions. The S-duality invariant fixed point is located at $\tau = i$ with $k_{mm} = k_{ee}$ and $k_{em}=0$. They are fixed by demanding that \eqref{s1} and \eqref{s2} are invariant under the S-duality (i.e. $a=d=0$, $b=1$,$c=-1$). Another interesting fixed point is the ST-duality invariant fixed point, which is located at $\tau = e^{-\frac{2\pi i}{3}}$ with $k_{ee} = k_{mm} = 2 k_{em}$.  They are fixed by demanding that \eqref{s1} and \eqref{s2} are invariant under the ST-duality (i.e. $a=0$, $b=1$,$c=-1$, $d=-1$). We may not trust our beta functions at the strong coupling as in the duality invariant fixed point, but regardless of the precise form of the beta functions, it is highly expected that these fixed points do exist. Toward the end of this section, we will see a concrete example of the fixed point of the latter type.}

Let us now come back to the equations of motion
\begin{align}
\frac{\mathrm{Im}(\tau)}{4\pi} \partial_\mu(F^{\mu\nu} + i *F^{\mu\nu}) = J_e^\nu + \tau J_m^\nu 
\end{align}
and consequential conservation laws $\partial^\mu J_e^\mu = \partial^\mu J_m^\mu = 0$. By unitarity, $F^{\mu\nu}$ and $\tilde{F}^{\mu\nu}$ acquires non-zero anomalous dimensions because otherwise we have $J_e^\mu = J_m^\mu =0$. At the same time, $J_e^\mu$ and $J_m^\mu$ also acquire anomalous dimensions and they are no longer primary operators. This is consistent only if we have no charged operators under $J_e^\mu$ and $J_m^\mu$ at the conformal fixed point. 
Indeed this should be the case once we gauge the electric as well as magnetic current: the only $U(1)_e\times U(1)_m$ neutral operators should be gauge invariant and physical. In this way, the resulting conformal field theory does not possess $U(1)_e \times U(1)_m$ global symmetry. 

While we have focused on the non-trivial conformal fixed point, we would like to point out one interesting thing about the renormalization group flow. Naively, one may think that the theory is obtained from a free Maxwell theory and $U(1)_e\times U(1)_m$ symmetric conformal field theory and add the ``relevant" gauging interactions. Then we might expect that the resultant fixed point theory would have smaller degrees of freedom (e.g. counted by the Weyl anomaly coefficient ``$a$") under the renormalization group flow \cite{Komargodski:2011vj}. However, because gauging electrically and magnetically is not a perturbation even at the ultraviolet, the situation is what we called ``strong gauging" in \cite{Nakayama:2015zaw}. Accordingly, there is no apparent reason why the Weyl anomaly coefficient must decrease along with this procedure. In other words, there is nowhere in the renormalization group flow where we have the decoupled Maxwell theory and the decoupled $U(1)_e\times U(1)_m$ symmetric conformal field theory because we cannot make the electric coupling and magnetic coupling simultaneously small.\footnote{We do not claim that it is impossible. From the conformal representation theory, it seems possible to have a recombination limit, where $U(1)$ gauge theory becomes free and the $U(1)_e\times U(1)_m$ global symmetry emerges. In order to realize such a scenario, we have to find more non-trivial matter and interactions than just simple gauging.} This should be contrasted with weakly gauging of conformal field theories (where we know that Weyl anomaly coefficient ``$a$" decreases). 

Let us take a look at one suggestive example. Suppose we have $\mathcal{N}=2$ supersymmetry and we have three free hyper multiplets. Under the $U(1)_e\times U(1)_m$ symmetry we would like to electrically and magnetically gauge, they have charge $(1,0)$, $(0,1)$ and $(1,1)$, satisfying the Dirac quantization condition. Accordingly, we have $k_{ee}=2$, $k_{em}= 1$ and $k_{mm}=2$. Following our prescription, we want to gauge the three hypermultiplets with one $U(1)$ vector multiplet electrically and magnetically. The resulting theory has a non-trivial renormalization group fixed point at $\tau = e^{-\frac{2\pi i}{3}}$. This model is believed to be an effective description of the simplest Argyres-Douglas superconformal field theory  \cite{Argyres:1995jj}.

The Euler Weyl anomaly coefficient of one vector multiplet and three hyper multiplet is given by $a = \frac{5}{24} + 3\frac{1}{24}= \frac{1}{3}$. On the other hand, the Euler Weyl anomaly coefficient of the Argyres-Douglas fixed point is $a=\frac{43}{120}>\frac{1}{3}$ \cite{Shapere:2008zf}. Thus, electrically and magnetically gauging seems to give a larger Euler Weyl anomaly coefficient. As we have mentioned, this is not inconsistent with the $a$-theorem which claims that the Euler Weyl anomaly coefficient should decrease along with the renormaliztaion group flow.

{As in the Argyres-Douglas example above, it is not difficult to generalize our idea in the $\mathcal{N}=1$ (or $\mathcal{N}=2$) supersymmetric setup. If we consider the supersymmetric QED with a massless hypermultiplet that has electric and magnetic charge $(q_e,q_m)$, the contribution to the holomorphic beta function is\footnote{Since the adjoint representation has zero electric or magnetic charge in $U(1)$ gauge theories, there is no difference between $\mathcal{N}=2$ and $\mathcal{N}=1$ supersymmetry here.}
\begin{align*}
\beta_\tau = -\frac{i}{2\pi} (q_m \tau + q_e )^2 \ .    
\end{align*}
This is consistent with our formula \eqref{betaf} (by noting we have extra contributions from complex scalars in addition to fermions), but is one-loop exact. }

\section{Discussions}

In this paper, we have proposed a new prescription to construct non-trivial conformal field theories by gauging electric and magnetic currents. The resulting theory is strongly coupled and it would be important to develop non-perturbative methods to study their properties. For instance, one may attempt the lattice regularization of Zwanziger action (see e.g. \cite{Cardy:1981qy}\cite{Barrie:2016wxf}) and take the thermodynamic limit.

Since the non-trivial renormalization group fixed point is infrared attractive (if the ungauged theory has no relevant deformations), we do not have to take the continuum limit. Indeed, it is very likely we cannot take the continuum limit unless we are exactly at the fixed point of the coupling constants (essentially due to the Landau-pole problem), but for our study of the long-distance physics as a conformal field theory, this tuning may not be necessary. Note that the lattice model \cite{Cardy:1981qy} seems to show a gapped phase (unless anomaly forbids it), but this is in part caused by the existence of relevant deformations (e.g. mass and interaction terms in the first quantized world-line matters).\footnote{Some recent developments of this model can be found in \cite{Honda:2020txe}\cite{Hayashi:2022fkw}. It is conjectured that if we increase the matter degrees of freedome, it becomes critical \cite{Anosova:2022yqx}.}

One direction is to study the constraint on the operator dimensions by using (numerical) conformal bootstrap (e.g. \cite{Poland:2018epd} for a review). One characteristic feature of our theory is the field strength $F_{\mu\nu}$ and $\tilde{F}_{\mu\nu}$ with anomalous dimensions. It would be interesting to study the bound on the operator spectrum appearing in $F_{\mu\nu}(x) \times F_{\rho\sigma}(0)$ from the conformal bootstrap.  For instance, the bound on the conformal dimension of a scalar operator should be non-trivial, and it might give a hint at a non-trivial conformal fixed point.\footnote{We, however, do know that if we consider a free Weyl fermion and construct the two-form operator $F_{\alpha\beta} = \lambda_{(\alpha} \lambda_{\beta)}$, then the operator product expansion gives a dimension $8$ scalar operator (rather than dimension $6$ operator) from the Pauli exclusion principle. As studied in \cite{Nakayama:2019jvm}, this  free example (rather than a non-trivial fixed point) may give a possible saturation of the bound.}

The other direction is to use holography. In the bottom-up approach, we may study the bulk effective field theories containing two-form tensor fields \cite{KN}. The anomalous dimension of the two-form operator is associated with Higgsing two-form tensor fields in the bulk. 
It would be interesting to see if this bottom-up construction can be embedded in string theory with the interpretation of electric/magnetic gauging.

In particle phenomenology,  $U(1)$ gauge theories typically suffer the Landau pole problem, and this is why we want to embed them in non-Abelian gauge theories. If it is allowed to introduce a magnetic coupling to a monopole, we see that the Landau pole problem is easily circumvented without such an embedding. It would be an interesting question if the Landau pole problem in non-Abelian gauge theories (e.g. due to too many flavors) can be circumvented in a similar manner. We also see new developments to compute scattering amplitude with monopoles \cite{Csaki:2020inw}\cite{Terning:2020dzg}\cite{Kitano:2021pwt}.

We would like to finish this paper with the following speculative idea. As we saw, if the $U(1)$ gauge theory contains electric as well as magnetic charges, the $\theta$-term is renormalized and it appears in the Weyl anomaly. Let us now consider the (gauged) supergravity. Typically, we only gauge the $U(1)_R$ current electrically. However, one may wonder what could happen if we gauge the $U(1)_R$ current electrically as well as magnetically.  We would obtain the $\theta$-term for the $U(1)_R$ gauge field in the Weyl anomaly. Then, the supersymmetry demands that it should accompany the Pontryagin term in the Weyl anomaly as well \cite{Nakagawa:2020gqc}. This mechanism could be used to find novel examples that show CP violating-terms in the Weyl anomaly \cite{Nakayama:2012gu}\cite{Nakayama:2018dig}\cite{Nakayama:2020cef}. 

\section*{Acknowledgements}
This work is in part supported by JSPS KAKENHI Grant Number 17K14301 and  Grant No. 21K03581. The author would like to thank Y.~Tanizaki for correspondence.

\end{document}